\documentclass{relaxed_system_lab}

\usepackage[T1]{fontenc}
\usepackage{charter}

\usepackage{amsmath,amssymb}

\usepackage{algorithm}
\usepackage{algorithmic}

\usepackage{enumitem}
\setlist[itemize]{topsep=2pt,itemsep=1pt,parsep=0pt,partopsep=0pt,leftmargin=1em}

\usepackage{xspace}

\newcommand{\sysname}{\texttt{V3DB}\xspace}
\newcommand{\comm}{\ensuremath{\mathsf{com}}\xspace}
\newcommand{\Verify}{\mathsf{Verify}}
\newcommand{\accept}{\mathtt{1}}
\newcommand{\reject}{\mathtt{0}}

\renewcommand{\paragraph}[1]{\noindent\textbf{#1}.}

\title{V3DB: Audit-on-Demand Zero-Knowledge Proofs for Verifiable Vector Search over Committed Snapshots}

\author[1,*]{Zipeng Qiu}
\author[2,*]{Wenjie Qu}
\author[2]{Jiaheng Zhang}
\author[1,\dagger]{Binhang Yuan}

\affiliation[1]{Hong Kong University of Science and Technology}
\affiliation[2]{National University of Singapore}

\contribution[*]{Equal contribution}
\contribution[\dagger]{Corresponding author}

\abstract{
Dense retrieval services increasingly underpin semantic search, recommendation, and retrieval-augmented generation, yet clients typically receive only a top-$k$ list with no auditable evidence of how it was produced.
We present \sysname, a verifiable, versioned vector-search service that enables audit-on-demand correctness checks for approximate nearest-neighbour (ANN) retrieval executed by a potentially untrusted service provider.
\sysname commits to each corpus snapshot and standardises an IVF-PQ search pipeline into a fixed-shape, five-step query semantics.
Given a public snapshot commitment and a query embedding, the service returns the top-$k$ payloads and, when challenged, produces a succinct zero-knowledge proof that the output is exactly the result of executing the published semantics on the committed snapshot---without revealing the embedding corpus or private index contents.
To make proving practical, \sysname avoids costly in-circuit sorting and random access by combining multiset equality/inclusion checks with lightweight boundary conditions.
Our prototype implementation based on Plonky2 achieves up to $22\times$ faster proving and up to $40\%$ lower peak memory consumption than the circuit-only baseline, with millisecond-level verification time.\\

Github Repo at \url{https://github.com/TabibitoQZP/zk-IVF-PQ}.
}

\begin{document}
\maketitle

\section{Introduction}

\begin{figure*}[t]
	    \centering
	    \includegraphics[width=0.99\linewidth]{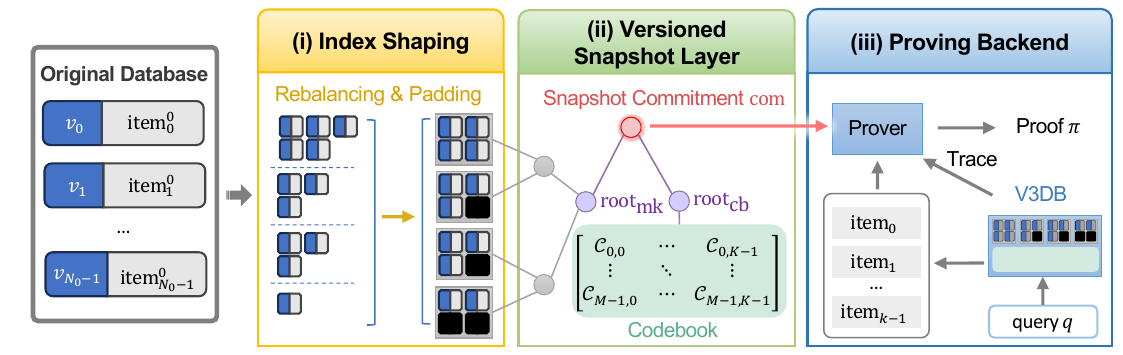}
	    \caption{Workflow of \sysname. (i) \emph{Index shaping}: from the original database $\{(v_t,\mathrm{item}^0_t)\}_{t=0}^{N_0-1}$, build a fixed-shape IVF-PQ snapshot via rebalancing and padding. (ii) \emph{Versioned snapshot layer}: publish a snapshot identifier $\comm=(\mathrm{root}_{\mathrm{mk}},\mathrm{root}_{\mathrm{cb}})$, where $\mathrm{root}_{\mathrm{mk}}$ is a Merkle root committing to the fixed-shape IVF layout and $\mathrm{root}_{\mathrm{cb}}$ is a hash digest of the PQ codebooks. (iii) \emph{Proving backend}: on query embedding $q$, return top-$k$ payloads $(\mathrm{item}_0,\dots,\mathrm{item}_{k-1})$ and, upon challenge, a succinct ZK proof $\pi$ that the list equals the output of the fixed-shape IVF-PQ semantics on the snapshot committed by $\comm$, while hiding snapshot contents and the trace.}
	    \label{fig:intro-overview}
\end{figure*}

Dense retrieval, i.e., similarity search over embedding vectors, has become a standard primitive for various applications such as semantic search, recommendation, and retrieval-augmented generation (RAG) \cite{lewis2020retrieval,zhong2024memorybank,qin2023toolllm,zhang2023repocoder}. This shift has driven the widespread adoption of vector-search services for approximate nearest-neighbour (ANN) queries over large embedding corpora \cite{jegou2010product,malkov2018efficient,johnson2019billion}. In many real-world deployments, the retrieval pipeline is executed by a remote service provider: a vector-search service maintains the embedding corpus and ANN index and exposes a retrieval API, while clients observe only the returned top-$k$ list. This separation creates a fundamental accountability gap: a retrieval API typically returns only the final ranked list, with little or no evidence of how the search was executed. When a client disputes a result, the service has no efficient way to demonstrate that it executes the mutually agreed-upon ANN procedure---using the promised parameters on the claimed snapshot of its corpus and index. In particular, the service could stealthily reduce the search budget, skip expensive steps, replay a stale snapshot, or introduce subtle ranking biases under operational or business pressure, and such deviations are difficult to detect from the top-$k$ output alone.
This raises the essential question: \emph{when challenged, can a retrieval service provide a short, efficiently checkable certificate that its returned top-$k$ list is exactly the output of a publicly specified ANN query semantics on a specific committed corpus snapshot, without requiring the client to download the corpus or rerun search from scratch?}

The ability to audit disputed retrieval results is important in high-value content settings such as patents,  finance, and legal search, where retrieval outputs can directly influence key decisions.
Commercial offerings increasingly incorporate semantic search over proprietary corpora (e.g., AI-powered patent semantic search and AI patent search in patent analytics platforms~\cite{questel_sophia_search,clarivate_derwent_ai_search}) and expose content-retrieval workflows via APIs~\cite{factiva_news_search_api,lexis_api}.
Moreover, a provider may have incentives to manipulate results in ways that are difficult to diagnose externally---for example, suppressing unfavourable items about a sponsor, boosting favourable coverage, or answering from an older snapshot that omits recent updates. At the same time, the underlying corpora, embeddings, and index structures are often proprietary, licensed, or sensitive (e.g., paywalled archives, internal repositories used for RAG, user preference profiles, or biometric templates). Releasing the corpus or index to ``prove innocence'' is therefore unacceptable. A practical audit mechanism must be compact and cheap to check, enabling \textit{audit-on-demand} dispute resolution by clients and third-party auditors, while disclosing only what the application intends to reveal.

Achieving private, auditable retrieval at a realistic scale is challenging because of the integrity and privacy requirements.
There are two broad approaches. One can run retrieval inside trusted execution environments (TEEs) and use remote attestation to provide integrity guarantees about a measured implementation, which can reduce disclosure but shifts trust to hardware vendors, attestation infrastructure, and side-channel assumptions; it also requires always-on protected execution, regardless of whether a query is ever audited later~\cite{sabt2015trusted,costan2016intel,costan2016sanctum}.
Alternatively, one can use cryptographic proofs that are generated on demand and can be checked by any party without trusting the operator. We focus on the latter, and in particular on succinct zero-knowledge (ZK) proofs to keep the corpus and index private during verification~\cite{blum2019non,parno2016pinocchio,groth16,ben2014succinct}.
At a high level, ZK proofs establish that there exists a private witness---including intermediate values of the retrieval execution---that satisfies a fixed set of arithmetic constraints; prover cost is therefore driven by the circuit size.

ZK proofs impose two practical constraints that shape verifiable retrieval.
\underline{First}, generating a ZK proof is typically substantially more expensive than plaintext execution and scales poorly with the number of constraints, so the proved computation must have a tightly controlled complexity budget.
\underline{Second}, the proved computation must be expressed as a fixed arithmetic circuit: data-dependent control flow (branching) and irregular memory access must be compiled into static relations, often making such steps expensive.
Consequently, certifying an \emph{exact} full scan over all $N$ $D$-dimensional vectors would require encoding $\Theta(ND)$ distance computations and would lead to an impractically large circuit.

This motivates us to focus on an efficient ANN pipeline that avoids full scans and admits a standardized, fixed-shape query procedure.
We therefore target inverted-file indices using product quantization (IVF-PQ)~\cite{jegou2010product}, whose query execution proceeds in a small sequence of budgeted steps.
However, even IVF-PQ contains primitives that are expensive in ZK circuits: probe selection and final top-$k$ extraction are comparison-heavy, candidate scoring relies on lookup tables indexed by PQ codes, and binding accessed snapshot values to a specific version requires authentication for soundness (e.g., Merkle paths inside the proof)~\cite{merkle1987digital}.

To the best of our knowledge, existing verifiable ANN query schemes in outsourced databases primarily rely on authenticated indices and verification objects (VOs). Such approaches can certify soundness and completeness but typically reveal access-dependent metadata, which is undesirable in many practical settings for large-scale dense retrieval over sensitive corpora~\cite{wang2024verifiable,cui2023towards}.

We address this integrity--privacy gap by instantiating retrieval evidence with succinct ZK proofs and present \sysname, a \emph{verifiable, versioned vector database} for dense retrieval at scale.
In \sysname, we standardize IVF-PQ into a fixed-shape, five-step query semantics that is amenable to proof generation.
Given a public snapshot commitment $\comm$ and a query embedding, the service returns a top-$k$ list of application-defined payloads and, upon challenge from the client, produces a succinct ZK proof certifying that the list is exactly the output of executing this semantics on the snapshot identified by $\comm$, while hiding the underlying corpus embeddings and private index contents from the verifier.

Figure~\ref{fig:intro-overview} sketches the end-to-end workflow of \sysname: the operator shapes a raw vector database into a fixed-shape IVF-PQ snapshot, commits to each snapshot version, and supports audit-on-demand ZK proofs for disputed query results.
We implement both a circuit-only baseline and our optimized multiset-based instantiation in Plonky2 and evaluate their end-to-end costs in detail; the multiset-based design reduces proving time by up to $\sim 22\times$ and peak memory by up to $\sim 40\%$ while keeping verification in milliseconds.

In summary, we make the following contributions:
\begin{itemize}
	\item \textbf{System and IVF-PQ instantiation.} We present \sysname for verifiable dense retrieval over committed snapshots. We instantiate it for IVF-PQ by standardizing a fixed-shape, five-step query semantics and designing a ZK-friendly index-shaping pipeline that enables fixed-shape execution at scale.
	\item \textbf{Efficient proving backend.} Instead of an impractical circuit-only baseline that directly encodes sorting/selection and random access in-circuit, we avoid these bottlenecks via multiset equality/inclusion checks plus lightweight boundary conditions, and analyze the resulting gate complexity.
    \item \textbf{Implementation and evaluation.} We implement both a baseline and an optimized design, and show through end-to-end evaluation that the optimized design makes audit-on-demand proof generation practical under realistic budgets.
\end{itemize}

\section{Preliminaries}

\subsection{ZK Proofs}
\label{subsec:pre-circuits}

Zero-knowledge succinct non-interactive arguments of knowledge (ZK-SNARKs) let a prover convince a verifier that a publicly stated computation is correct, without revealing private witness values beyond what is implied by the public inputs/outputs~\cite{goldwasser2019knowledge,blum2019non,parno2016pinocchio,groth16}.
In \sysname, the public interface consists of a query embedding $q$, a snapshot version commitment $\comm$, and a returned top-$k$ payload list that the verifier checks against the claimed computation.

\paragraph{Circuits as polynomial constraints}
In circuit-based SNARKs, the proved computation is expressed as a fixed constraint system: a set of low-degree algebraic relations over a field $\mathbb{F}$.
The prover supplies witness values for intermediate wires, and the verifier checks a short proof that these constraints are satisfiable with the claimed public inputs alone.
As a minimal example, suppose a circuit computes two candidate values $x_0,x_1\in\mathbb{F}$ (e.g., from two alternative branches) and uses a private bit $b\in\{0,1\}$ to select the public output $y$ accordingly.
One can enforce:
$b(b-1)=0,\: y=(1-b)x_0+bx_1$.
This encodes conditional behavior without runtime branching.

\paragraph{Fixed-shape execution}
A circuit has a predetermined set of constraints and fixed loop bounds across all instances, so data-dependent control flow and irregular memory addressing must be compiled into static relations (e.g., algebraic selection and fixed-size scans).
In the example above, if $x_0$ and $x_1$ come from two expensive branches, a plaintext program evaluates only one branch, but a circuit must constrain \emph{both} branches and then select using $b$.
This motivates our fixed-shape IVF-PQ semantics and our focus on avoiding costly in-circuit sorting/selection and random access.

\paragraph{Fixed-point representation and range bounds}
Circuit constraints are defined over a finite field $\mathbb{F}$, so we represent real-valued embeddings and centroids using fixed-point integers embedded in $\mathbb{F}$ under a public scaling and bit-width chosen per deployment.
Our proofs therefore attest to correctness with respect to this encoded representation in the circuit.
To make comparisons and distance computations well defined, we range-bound all relevant witness values so that intermediate results do not wrap around modulo $\mathbb{F}$.
Our experiments instantiate this encoding in our prototype and quantify its impact on retrieval utility in practice.

\paragraph{Gate counts and proving time (binning effect)}
We track circuit size by the number of gate rows $G$ in a PLONK-like system~\cite{plonk}.
Modern provers use FFT-based polynomial operations and pad to a power-of-two evaluation domain~\cite{ben2018fast}, so we define $G_B := 2^{\lceil\log_2 G\rceil}$ and approximate proving time as $\Theta(G_B\log_2 G_B)$.
As a result, reducing $G$ can yield outsized speedups when it pushes the circuit into a smaller $G_B$ bin.
For the example constraints above, treating each add/sub/mul as $\Theta(1)$ rows gives a crude estimate $G\approx 6$ (1~mul + 1~sub for the booleanity constraint and 2~mul + 1~sub + 1~add for the selection constraint), hence $G_B=8$.

\subsection{IVF-PQ}
\label{subsec:ivf_pq_intro}

IVF-PQ is a widely used ANN index that avoids a full scan by probing a small number of inverted lists and using compact PQ codes for fast approximate \emph{scoring}~\cite{jegou2010product,johnson2019billion}.

\paragraph{Coarse partition (IVF)}
IVF first trains a set of coarse centroids (e.g., via $k$-means~\cite{lloyd1982least,arthur2006k}) and assigns each database vector to its nearest centroid, forming one inverted list per centroid.
At query time, instead of scanning the database, the search probes only several of the closest inverted lists, yielding a controlled scan budget.

\paragraph{Product quantization (PQ) and table-based scoring}
To reduce storage and speed up per-candidate scoring, IVF-PQ applies product quantization to database residuals (database vector minus centroid).
Each database vector is stored as a code: the residual is split into blocks, each block is quantized by a small codebook, and the resulting code indices are stored.
At query time, the engine builds asymmetric distance computation (ADC) lookup tables from the query residual for each probed list and uses the PQ code indices to fetch and sum table entries, producing approximate distances.

\paragraph{From plain IVF-PQ to verifiable semantics}
Together, the IVF and PQ components make IVF-PQ a natural target for audit-on-demand verification: query processing decomposes into a small sequence of steps with explicit budgets (probing and scanning), enabling a well-scoped correctness statement.
However, directly proving the plaintext pipeline in a circuit is inefficient.
First, inverted lists have variable lengths, inducing data-dependent loop bounds and memory accesses, while circuits require fixed-shape execution.
Second, probe selection and final top-$k$ extraction rely on ordering, and PQ scoring uses data-dependent table lookups; both translate to comparison-heavy selection and random access in circuits.
To obtain a circuit-friendly statement, \sysname standardizes IVF-PQ into a fixed-shape semantics by enforcing a per-list capacity bound via rebalancing and padding each list to a uniform number of slots with a validity flag.
We detail index shaping and formalize the five-step semantics in Section~\ref{sec:circuit_design}.

\paragraph{Notation and terminology}
For the remainder of the paper, we use the following notation.
Let $n_{\mathrm{list}}$ denote the number of inverted lists (coarse centroids) and write the centroid table as $\boldsymbol{\mu}:=(\mu_0,\dots,\mu_{n_{\mathrm{list}}-1})$, where $\mu_i$ is the centroid associated with list $i$.
An \emph{IVF layout} consists of this centroid table together with the corresponding inverted lists.
For PQ, we use $M$ sub-quantizers, each with a codebook of size $K$; we write the \emph{PQ codebooks} as $\{\mathcal{C}_{m,k}\}$, where $m\in\{0,\dots,M-1\}$ and $k\in\{0,\dots,K-1\}$.
Intuitively, each $\mathcal{C}_{m,k}$ is a learned codeword used for table-based scoring.
In \sysname, we work with a \emph{fixed-shape IVF layout} obtained by rebalancing and padding each list to exactly $n$ slots (a public per-list capacity bound) with an explicit validity flag, and we refer to the fixed-shape IVF layout together with the PQ codebooks as a \emph{snapshot}.

\section{System Overview}
\label{sec:overview}


\subsection{Workflow and Problem Statement}
\label{subsubsec:problem_formulation}

This subsection introduces \sysname in four parts.
The first three are the workflow components depicted in Figure~\ref{fig:intro-overview}: index shaping, the versioned snapshot commitment layer, and the proving backend.
We then specify the public interface and the proved statement that the verifier checks.
Throughout, we fix a public IVF-PQ configuration, which determines both the plaintext workflow and the circuit shape; Table~\ref{tab:ivfpq_params} summarizes the notation used throughout the paper.

\begin{table}[t]
\centering
\small
\setlength{\tabcolsep}{6pt}
\begin{tabular}{ll}
\toprule
Symbol & Meaning \\
\midrule
$N_0$ & number of vectors in the dataset (before shaping/padding) \\
$D$ & embedding vector dimension \\
$n_{\mathrm{list}}$ & number of inverted lists (coarse centroids) \\
$n_{\mathrm{probe}}$ & number of lists probed per query \\
$n$ & per-list capacity bound (each list padded to exactly $n$ slots) \\
$M$ & number of PQ sub-quantizers \\
$K$ & codebook size per sub-quantizer \\
$k$ & top-$k$ payload list size \\
$t_{\mathrm{cmp}}$ & comparison bit-length parameter in circuits \\
\bottomrule
\end{tabular}
\caption{Key notation for our fixed-shape IVF-PQ configuration, used throughout the paper.}
\label{tab:ivfpq_params}
\end{table}

\paragraph{Index shaping (offline)}
Given a raw dataset $\{(v_t,\mathrm{item}^0_t)\}_{t=0}^{N_0-1}$ of $N_0$ embedding vectors $v_t$ with application-defined payloads $\mathrm{item}^0_t$ and a public configuration (Table~\ref{tab:ivfpq_params}), the operator builds a fixed-shape IVF-PQ snapshot by applying capacity-constrained rebalancing so that each list has at most $n$ valid records and then padding to exactly $n$ slots with a validity flag per slot.

\paragraph{Versioned snapshot layer (offline)}
The operator commits to each snapshot version by publishing a public version identifier $\comm$ before serving queries.
We detail the commitment construction and snapshot-binding constraints in Section~\ref{sec:circuit_design}.
When the database is updated, the operator repeats index shaping and publishes a new commitment, yielding an epoch-style version history.

\paragraph{Proving backend (online/audit)}
Given a query embedding $q$ and a claimed snapshot identifier $\comm$, the service returns a top-$k$ payload list $(\mathrm{item}_0,\dots,\mathrm{item}_{k-1})$.
Upon challenge, it produces a succinct proof $\pi$ certifying that there exists a private snapshot consistent with $\comm$ such that the returned list is exactly the output of our standardized five-step, fixed-shape IVF-PQ semantics.
The ZK property hides the snapshot contents and the execution trace from the verifier during verification.

\paragraph{Public interface and proved statement}
Given $(\comm,q)$, the service returns $(\mathrm{item}_0,\dots,\mathrm{item}_{k-1})$ to the verifier.
Upon challenge, it produces a proof $\pi$ such that
\[
\Verify\bigl(\comm,q,(\mathrm{item}_0,\dots,\mathrm{item}_{k-1}),\pi\bigr)\in\{\accept,\reject\}.
\]
The verifier accepts only when the returned list exactly matches the output of the standardized semantics on the snapshot version committed by $\comm$.
The only disclosed information is what the application reveals through $(\comm,q)$ and the returned list.

\subsection{Threat Model}
\label{subsubsec:threat_model}

We focus on the integrity of outsourced dense retrieval.
We consider a malicious retrieval operator who controls the corpus, index, and execution and may deviate arbitrarily from the publicly specified workflow while still returning a syntactically valid top-$k$ list.
We target integrity violations, including (i) \emph{workflow deviation} (e.g., downgrading parameters, skipping steps, cherry-picking candidates, or biasing ranking) and (ii) \emph{version equivocation} (e.g., answering using a stale or mixed snapshot while claiming a committed version).
The soundness of proof statement rules out \emph{any} deviation that would make the returned list differ from the standardized semantics on a snapshot consistent with the claimed commitment $\comm$.

\paragraph{Scope}
We treat $q$ as a public embedding and focus on proving the retrieval stage.
Proving the embedding computation (e.g., the neural inference that produces $q$) is orthogonal and has been studied by recent ZKML systems~\cite{zkml,feng2024zeno}.
We also do not aim to provide access-pattern privacy or prevent denial-of-service (e.g., refusing to answer or refusing to generate a proof); rather, we make incorrect execution detectable whenever a proof is produced.
Finally, \sysname proves correctness relative to a public commitment $\comm$; selecting which commitment is current/authorized is handled by the application layer (e.g., signed bulletins or transparency logs).

\subsection{Proving Cost Overview}
\label{subsec:overview_efficiency}

\begin{table*}[t]
\centering
\small
\setlength{\tabcolsep}{6pt}
\begin{tabular}{lccc}
\toprule
Step & Circuit-only baseline & Multiset instantiation & Key idea \\
\midrule
(i) Centroid distances & $\Theta(n_{\mathrm{list}} D)$ & same & --- \\
(ii) Probe selection & $\Theta(n_{\mathrm{probe}}n_{\mathrm{list}}t_{\mathrm{cmp}})$ & $\Theta(n_{\mathrm{list}}t_{\mathrm{cmp}})$ & multiset equality + boundary checks \\
(iii) ADC lookup tables & $\Theta(n_{\mathrm{probe}}KD)$ & same & --- \\
(iv) Candidate distances & $\Theta(n_{\mathrm{probe}}nMK)$ & $\Theta(t_{\mathrm{cmp}}n_{\mathrm{probe}}M\max\{K,n\})$ & multiset inclusion + comparisons \\
(v) Final Top-$k$ selection & $\Theta(kn_{\mathrm{probe}}nt_{\mathrm{cmp}})$ & $\Theta(n_{\mathrm{probe}}nt_{\mathrm{cmp}})$ & multiset equality + boundary checks \\
\bottomrule
\end{tabular}
\caption{Dominant asymptotic arithmetic-gate counts of the query-semantics circuit per standardized IVF-PQ step, comparing the circuit-only baseline and our multiset-based instantiation. Notation follows Table~\ref{tab:ivfpq_params}. Snapshot binding adds an additional $\Theta(n_{\mathrm{probe}}\log n_{\mathrm{list}})$ Merkle-opening overhead (Equation~\ref{eq:gcommit}).}
\label{tab:overview_costs}
\end{table*}

In \sysname, the dominant audit-on-demand cost is proof generation rather than plaintext query execution, so we focus on minimizing the gate count needed to express our standardized semantics.
A direct circuit-only encoding is expensive because IVF-PQ relies heavily on (i) ordering operations for probe selection and final top-$k$ extraction and (ii) data-dependent table lookups for PQ-based scoring.
The circuit-only baseline encodes these primitives directly in-circuit, while our multiset-based design shifts them off-circuit and proves lightweight consistency relations (multiset equality/inclusion plus boundary checks).

\paragraph{Budget abstractions}
To discuss configuration trade-offs, we use two standard IVF-PQ tuning notions~\cite{jegou2010product}.
The \emph{scan budget} is the number of PQ codes evaluated per query,
\[
N_{\mathrm{sel}}:=n_{\mathrm{probe}}\cdot n,
\]
and the \emph{code budget} is the number of bits stored per vector,
\[
B:=M\log_2 K,
\]
where we assume $K$ is a power of two as in standard IVF-PQ deployments.
When the padded capacity is fixed to $N:=n_{\mathrm{list}}\cdot n$, fixing $N_{\mathrm{sel}}$ is equivalent to fixing the probing ratio $r:=n_{\mathrm{probe}}/n_{\mathrm{list}}=N_{\mathrm{sel}}/N$, since $N_{\mathrm{sel}}=r\cdot N$.
Accordingly, later references to a ``fixed scan budget'' may specify either $N_{\mathrm{sel}}$ or $(N,r)$, depending on context.
We use these notions to express both our cost-driven configuration selection and the evaluation sweeps.

\paragraph{Five-step semantics and per-step costs}
Under our standardized semantics, query execution follows five steps:
(i) compute centroid distances;
(ii) select the $n_{\mathrm{probe}}$ closest lists;
(iii) construct ADC lookup tables for the probed lists;
(iv) score all candidate slots via PQ-table lookups; and
(v) extract the final top-$k$ payloads.
Table~\ref{tab:overview_costs} summarizes the dominant gate-count terms of each step under the circuit-only baseline and our multiset-based instantiation.

Summing the dominant terms yields the gate complexity:
\begin{align}
G_{\mathrm{baseline}} &= \Theta\Bigl(n_{\mathrm{list}}D + n_{\mathrm{probe}}n_{\mathrm{list}}t_{\mathrm{cmp}} + n_{\mathrm{probe}}KD \notag\\
&\hspace{2.2em} + n_{\mathrm{probe}}nMK + kn_{\mathrm{probe}}nt_{\mathrm{cmp}}\Bigr), \label{eq:gbaseline}\\
G_{\mathrm{multiset}} &= \Theta\Bigl(n_{\mathrm{list}}D + n_{\mathrm{list}}t_{\mathrm{cmp}} + n_{\mathrm{probe}}KD \notag\\
&\hspace{2.2em} + t_{\mathrm{cmp}}n_{\mathrm{probe}}M\max\{K,n\} + n_{\mathrm{probe}}nt_{\mathrm{cmp}}\Bigr). \label{eq:gmultiset}
\end{align}
Let $G_{\mathrm{commit}}$ denote the gate count of the snapshot commitment-check circuit. Under a hash-cost abstraction, our analysis yields
\begin{equation}
G_{\mathrm{commit}}=\Theta\Bigl(KD + n_{\mathrm{list}}D + n_{\mathrm{probe}}(nM + D + \log n_{\mathrm{list}})\Bigr).
\label{eq:gcommit}
\end{equation}
The end-to-end proof circuit composes the query-semantics constraints (Equations~\ref{eq:gbaseline}--\ref{eq:gmultiset}) with the snapshot-binding constraints (Equation~\ref{eq:gcommit}).
This yields the dominant end-to-end gate complexity:
\begin{align}
G^{\mathrm{total}}_{\mathrm{baseline}} &= \Theta\Bigl(n_{\mathrm{list}}D + n_{\mathrm{probe}}n_{\mathrm{list}}t_{\mathrm{cmp}} + n_{\mathrm{probe}}KD \notag\\
&\hspace{2.2em} + n_{\mathrm{probe}}nMK + kn_{\mathrm{probe}}nt_{\mathrm{cmp}} \notag\\
&\hspace{2.2em} + n_{\mathrm{probe}}\log n_{\mathrm{list}}\Bigr), \label{eq:gtotal-baseline}\\
G^{\mathrm{total}}_{\mathrm{multiset}} &= \Theta\Bigl(n_{\mathrm{list}}D + n_{\mathrm{list}}t_{\mathrm{cmp}} + n_{\mathrm{probe}}KD \notag\\
&\hspace{2.2em} + t_{\mathrm{cmp}}n_{\mathrm{probe}}M\max\{K,n\}\notag\\
&\hspace{2.2em} + n_{\mathrm{probe}}nt_{\mathrm{cmp}} + n_{\mathrm{probe}}\log n_{\mathrm{list}}\Bigr). \label{eq:gtotal-multiset}
\end{align}
Compared with the query-semantics costs $G_{\mathrm{baseline}}$ and $G_{\mathrm{multiset}}$, snapshot binding introduces an additional $\Theta(n_{\mathrm{probe}}\log n_{\mathrm{list}})$ Merkle-authentication term (Equation~\ref{eq:gcommit}).
Experiment~3 shows this overhead is small in our evaluated regimes; we therefore omit the $\Theta(n_{\mathrm{probe}}\log n_{\mathrm{list}})$ term in later scaling discussions unless stated otherwise.
We leverage these expressions to derive a configuration-selection procedure under fixed budgets (Section~\ref{subsec:config_selection}).

\section{Circuit Design}
\label{sec:circuit_design}

This section describes how \sysname realizes our standardized fixed-shape IVF-PQ semantics as succinct ZK circuits.
We prove the verifiable-retrieval statement from Section~\ref{subsubsec:problem_formulation}: given public inputs $(\comm, q, (\mathrm{item}_0,\dots,\mathrm{item}_{k-1}))$, the prover shows that the returned payload list is exactly the output of the reference semantics on the snapshot version committed by $\comm$.
The witness includes snapshot values (centroids, PQ codebooks, opened inverted-list records, and Merkle authentication paths) and intermediate trace values required by the semantics; all of these remain hidden by ZK, while verification remains public and cheap.

We construct two end-to-end circuit instantiations for the same statement.
The circuit-only baseline directly encodes ordering and random access in-circuit, while our multiset-based design shifts these expensive primitives off-circuit and verifies consistency using multiset equality/inclusion plus lightweight boundary checks.
Both instantiations prove the same public interface and differ only in how the reference relations are enforced.
We use the shorthand $[Z]:=\{0,1,\dots,Z-1\}$ to denote integer index ranges.

\subsection{Index Shaping via Rebalancing}
\label{subsubsec:adapt_rebalance}

After standard $k$-means, inverted lists can be highly imbalanced.
Let $\mathcal{X}_i$ denote the set of database vectors assigned to centroid $\mu_i$ and let $n_i:=|\mathcal{X}_i|$ be its size.
A naive fixed-shape layout sets the padded list capacity to the maximum cluster size,
\[
n=\max\{n_0,n_1,\dots,n_{n_{\mathrm{list}}-1}\},
\]
which is unstable: a single outlier cluster inflates padding for all lists and increases both the scan cost and the circuit size.

Instead, \sysname fixes a hard capacity bound $n$ in advance and post-processes the initial assignments using a lightweight rebalancing procedure inspired by~\cite{bradley2000constrained} to enforce a uniform per-list bound:
\[
|\mathcal{X}_i|\le n \qquad \text{for all } i\in[n_{\mathrm{list}}].
\]
With $N := n_{\mathrm{list}}\cdot n \ge N_0$, Algorithm~\ref{alg:rebalance} iteratively moves points out of overfull clusters to the nearest available underfull clusters until all clusters satisfy the bound.
We then pad each list to exactly $n$ slots and attach a validity flag $f_{i,j}\in\{0,1\}$ for each slot.
Each padded slot stores a record $(f_{i,j},\,\mathrm{item}_{i,j},\,v^{\mathrm{PQ}}_{i,j})$, where $\mathrm{item}_{i,j}$ is an application-defined payload and $v^{\mathrm{PQ}}_{i,j}$ is the PQ code; for padding slots ($f_{i,j}=0$), the remaining fields may be arbitrary.

\begin{algorithm}[htb]
\caption{Capacity-constrained cluster rebalancing to enforce the per-list bound $|\mathcal{X}_i|\le n$ in the fixed-shape IVF layout.}\label{alg:rebalance}
\begin{algorithmic}
\footnotesize
\REQUIRE Centroid table $\boldsymbol{\mu}=(\mu_0,\mu_1,\dots,\mu_{n_{\mathrm{list}}-1})$
\REQUIRE Clustered points $\mathcal{X}_i=\{v_{i,j}\mid j\in[n_i]\}$ for each cluster $i$
\REQUIRE Capacity bound $n$
\ENSURE Updated clusters $\{\mathcal{X}_i\}$ such that $|\mathcal{X}_i|\le n$ for all $i$

\WHILE{$\exists\, i \text{ s.t. } n_i > n$}
    \STATE $O \gets \{i \mid n_i > n\}$,\quad $F \gets \{t \mid n_t < n\}$ \COMMENT{\textcolor{blue}{overfull / free clusters}}
    \STATE $\mathcal{M} \gets \varnothing$ \COMMENT{\textcolor{blue}{candidate moves}}
    \FOR{each $i \in O$}
        \FOR{each $v \in \mathcal{X}_i$}
            \STATE $t^\star \gets \arg\min_{t \in F} \|v - \mu_t\|_2^2$
            \STATE $\Delta(v,i) \gets \|v - \mu_{t^\star}\|_2^2 - \|v - \mu_i\|_2^2$
            \STATE Add $(v, i, t^\star, \Delta(v,i))$ to $\mathcal{M}$
        \ENDFOR
    \ENDFOR
    \STATE Sort $\mathcal{M}$ by increasing $\Delta$
    \FOR{each $(v, i, t, \Delta) \in \mathcal{M}$ in order}
        \IF{$n_i > n$ \AND $n_t < n$ \AND $v \in \mathcal{X}_i$}
            \STATE $\mathcal{X}_i \gets \mathcal{X}_i \setminus \{v\}$,\quad $\mathcal{X}_t \gets \mathcal{X}_t \cup \{v\}$
            \STATE $n_i \gets n_i - 1$,\quad $n_t \gets n_t + 1$
        \ENDIF
    \ENDFOR
\ENDWHILE
\RETURN $\{\mathcal{X}_i\}_{i=0}^{n_{\mathrm{list}}-1}$
\end{algorithmic}
\end{algorithm}

\subsection{Fixed-Shape Five-Step IVF-PQ Semantics}
\label{subsubsec:adapt_query}

We specify the fixed-shape snapshot model used by our circuits and by the reference semantics.\label{subsec:ivf_pq_model}
The parameters are: database size $N_0$, padded capacity $N:=n_{\mathrm{list}}\cdot n$, vector dimension $D$, number of inverted lists $n_{\mathrm{list}}$, number of probed lists $n_{\mathrm{probe}}$, top-$k$ size $k$, per-list capacity $n$, and PQ parameters $(M,K)$ where $D=Md$ and each block uses a codebook of size $K$. The committed snapshot contains:
\begin{itemize}
    \item \textit{Centroids} $\boldsymbol{\mu}=(\mu_0,\dots,\mu_{n_{\mathrm{list}}-1})$ with $\mu_i\in\mathbb{F}^D$.
    \item \textit{PQ codebooks} $\{\mathcal{C}_{m,k}\}_{m\in[M],k\in[K]}$ with $\mathcal{C}_{m,k}\in\mathbb{F}^d$.
    \item \textit{Fixed-shape inverted lists.}
    For each list $i\in[n_{\mathrm{list}}]$ and slot $j\in[n]$, the snapshot stores a record
    $(f_{i,j},\mathrm{item}_{i,j},v^{\mathrm{PQ}}_{i,j})$, where $f_{i,j}\in\{0,1\}$ is a validity flag, $\mathrm{item}_{i,j}$ is an application-defined payload, and
    $v^{\mathrm{PQ}}_{i,j}=(v_{i,j}^{(0)},\dots,v_{i,j}^{(M-1)})\in[K]^M$ is the PQ code.
    If $f_{i,j}=0$, the slot is padding and the remaining fields are arbitrary.
\end{itemize}

We use squared $\ell_2$ distance over $\mathbb{F}$ throughout:
\[
\mathsf{dist}(x,y) := \sum_{t=0}^{D-1} (x_t-y_t)^2.
\]

We now specify the IVF-PQ query computation that our proof system attests to.
Given a query $q\in\mathbb{F}^D$ and a public snapshot commitment $\comm$, the prover claims that there exists a fixed-shape snapshot consistent with $\comm$ such that the public output is a length-$k$ list of payloads
$(\mathrm{item}_0,\dots,\mathrm{item}_{k-1})$ produced by the following five-step procedure for that query.

\paragraph{Step 1. Centroid distances}
Define centroid distances
\[
d_i := \mathsf{dist}(q,\mu_i), \qquad i\in[n_{\mathrm{list}}],
\]
and the index--distance sequence
\[
S^{\mathrm{orig}}_{\mathrm{idx,dist}}
:= \big((0,d_0),(1,d_1),\dots,(n_{\mathrm{list}}-1,d_{n_{\mathrm{list}}-1})\big).
\]

\paragraph{Step 2. Probe selection}
Sort $S^{\mathrm{orig}}_{\mathrm{idx,dist}}$ by the distance to obtain
\[
S^{\mathrm{sorted}}_{\mathrm{idx,dist}}
:= \big((i_0,d_{i_0}),(i_1,d_{i_1}),\dots,(i_{n_{\mathrm{list}}-1},d_{i_{n_{\mathrm{list}}-1}})\big),
\]
where $(i_0,\dots,i_{n_{\mathrm{list}}-1})$ is a permutation of $[n_{\mathrm{list}}]$.
We only require the first $n_{\mathrm{probe}}$ entries to be ordered and to have distances no larger than the remaining ones:
\[
\begin{aligned}
d_{i_0}\le d_{i_1}\le \cdots \le d_{i_{n_{\mathrm{probe}}-1}}
&\quad \text{and} \\
d_{i_{n_{\mathrm{probe}}-1}} \le d_{i_t}
&\quad \text{for all } t\in\{n_{\mathrm{probe}},\dots,n_{\mathrm{list}}-1\}.
\end{aligned}
\]
The probed set is defined as the first $n_{\mathrm{probe}}$ indices,
\[
P(q) := (i_0,\dots,i_{n_{\mathrm{probe}}-1}), \qquad |P(q)| = n_{\mathrm{probe}}.
\]

\paragraph{Step 3. ADC lookup tables}
For each probed list $i\in P(q)$, form the residual query $q-\mu_i$, split it into $M$ blocks
$(q-\mu_i)^{(m)}\in\mathbb{F}^d$ (recall $D=Md$), and define the ADC lookup tables
\[
\mathsf{LUT}_{i,m,k}
:= \mathsf{dist}\big(\mathcal{C}_{m,k}, (q-\mu_i)^{(m)}\big),
\qquad m\in[M],\ k\in[K].
\]

\paragraph{Step 4. Candidate distances}
For each $i\in P(q)$ and each slot $j\in[n]$, define the ADC distance
\[
\widetilde{d}_{i,j}
:= \sum_{m=0}^{M-1} \mathsf{LUT}_{i,m,\,v_{i,j}^{(m)}}.
\]
To ensure padded entries never affect the result, fix a public constant $d_{\max}$ that exceeds any attainable
$\widetilde{d}_{i,j}$ for valid slots (under the enforced range bounds), and define the masked distance
\[
D_{i,j} := f_{i,j}\cdot \widetilde{d}_{i,j} + (1-f_{i,j})\cdot d_{\max}.
\]
Collect all candidate pairs into the flattened sequence
\[
S^{\mathrm{orig}}_{\mathrm{item,dist}}
:= \big((\mathrm{item}_{i,j},D_{i,j}) \ \big|\  i\in P(q),\ j\in[n]\big),
\]
which has length $N_{\mathrm{sel}}$ (i.e, the scan budget).

\paragraph{Step 5. Final top-$k$ selection}
View $S^{\mathrm{orig}}_{\mathrm{item,dist}}$ as a length-$N_{\mathrm{sel}}$ sequence of pairs. Sort it by the distance component to obtain
\[
S^{\mathrm{sorted}}_{\mathrm{item,dist}}
:= \big((\mathrm{item}_0,D_0),(\mathrm{item}_1,D_1),\dots,(\mathrm{item}_{N_{\mathrm{sel}}-1},D_{N_{\mathrm{sel}}-1})\big),
\]
where we only require the first $k$ entries to be ordered and to have distances no larger than the remaining ones:
\[
\begin{aligned}
D_0\le D_1\le \cdots \le D_{k-1}
&\quad \text{and} \\
D_{k-1} \le D_t
&\quad \text{for all } t\in\{k,\dots,N_{\mathrm{sel}}-1\}.
\end{aligned}
\]
The query output is the first $k$ items in this order:
\[
(\mathrm{item}_0,\mathrm{item}_1,\dots,\mathrm{item}_{k-1}).
\]

\subsection{Snapshot Commitments}
\label{subsubsec:snapshot_commitment}

To bind each proof to a fixed IVF-PQ snapshot version while keeping the snapshot contents private, the operator publishes a public commitment $\comm$ before answering queries.
We instantiate $\comm$ using a SNARK-friendly hash function $\mathsf{Hash}$ and Merkle trees~\cite{merkle1987digital}.
In Plonky2, $\mathsf{Hash}$ is instantiated with Poseidon~\cite{poseidon}.
We model $\mathsf{Hash}$ as mapping a fixed-length sequence of field elements to a single digest in $\mathbb{F}$.

\begin{figure}[htbp]
    \centering
    \includegraphics[width=0.5\linewidth]{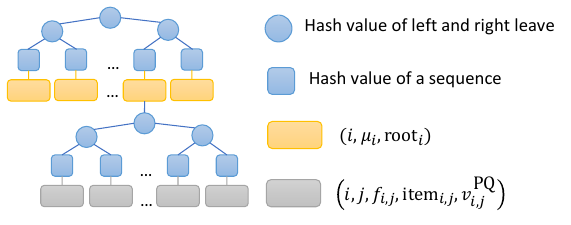}
    \caption{Hierarchical Merkle commitment for the fixed-shape IVF layout. The top-level root $\mathrm{root}_{\mathrm{mk}}$ binds the centroid table $\boldsymbol{\mu}$ and the padded inverted-list slot records.}
    \label{fig:merkle_commitment}
\end{figure}

We commit to an IVF-PQ snapshot in two parts: (i) a hierarchical Merkle root $\mathrm{root}_{\mathrm{mk}}$ covering the fixed-shape IVF layout (the centroid table $\boldsymbol{\mu}$ and padded inverted lists) and (ii) a hash digest $\mathrm{root}_{\mathrm{cb}}$ covering the PQ codebooks.
We write $\comm := (\mathrm{root}_{\mathrm{mk}},\mathrm{root}_{\mathrm{cb}})$.
Figure~\ref{fig:merkle_commitment} illustrates the resulting hierarchy of $\mathrm{root}_{\mathrm{mk}}$.

\paragraph{Inverted-list leaves and list roots}
For each list $i\in[n_{\mathrm{list}}]$ and slot $j\in[n]$, define the leaf digest
\[
\mathrm{hash}_{i,j} := \mathsf{Hash}\bigl(i,j,f_{i,j},\mathrm{item}_{i,j},v^{\mathrm{PQ}}_{i,j}\bigr),
\]
where $\mathsf{Hash}(\cdot)$ absorbs the $M$ components of $v^{\mathrm{PQ}}_{i,j}$ (from $v_{i,j}^{(0)}$ to $v_{i,j}^{(M-1)}$) in a fixed canonical order.
Let $\mathrm{root}_i$ denote the Merkle root of list $i$, computed as
\[
\mathrm{root}_i := \mathsf{MerkleTree}(\mathrm{hash}_{i,0},\dots,\mathrm{hash}_{i,n-1}).
\]
For simplicity, we assume $n$ is a power of two so the Merkle tree is full and balanced; otherwise, the snapshot layer can pad to the next power of two.

\paragraph{Centroid binding and top-level root}
We bind each centroid together with its list root by defining
\[
\mathrm{hash}_i := \mathsf{Hash}(i,\mu_i,\mathrm{root}_i),
\]
where $\mathsf{Hash}(i,\mu_i,\mathrm{root}_i)$ absorbs the $D$ coordinates of $\mu_i$ in a fixed canonical order.
The top-level commitment to the IVF layout is
\[
\mathrm{root}_{\mathrm{mk}} := \mathsf{MerkleTree}(\mathrm{hash}_0,\dots,\mathrm{hash}_{n_{\mathrm{list}}-1}).
\]
Similarly, we assume $n_{\mathrm{list}}$ is a power of two (or pad in the commitment layer when needed).

\paragraph{Codebook digest}
Since PQ codebooks are static once trained, we commit to them by hashing a canonical flattening:
\[
\mathrm{root}_{\mathrm{cb}} := \mathsf{Hash}(\mathcal{C}_{0,0},\dots,\mathcal{C}_{0,K-1},\mathcal{C}_{1,0},\dots,\mathcal{C}_{M-1,K-1}).
\]

\subsection{Snapshot-Binding Circuit}
\label{subsec:commitment_circuit}

\paragraph{Commitment-check constraints}
Our end-to-end statement binds the private snapshot values used in the five-step semantics to the public commitment $\comm$.
The query circuit uses witness values for: (i) the centroid table $\boldsymbol{\mu}$ used in Steps~1--2, (ii) the PQ codebooks $\{\mathcal{C}_{m,k}\}$ used in Step~3, and (iii) the padded slot records of the probed inverted lists used in Step~4.
The commitment-check subcircuit enforces three groups of constraints:

\noindent\textbf{Codebooks.}
Recompute $\mathrm{root}'_{\mathrm{cb}}$ by hashing the canonical flattening of the PQ codebooks and enforce $\mathrm{root}'_{\mathrm{cb}}=\mathrm{root}_{\mathrm{cb}}$.

\noindent\textbf{Centroids and list digests.}
Since probe selection depends on centroid distances to \emph{all} lists, the circuit binds all witness centroids to $\mathrm{root}_{\mathrm{mk}}$ together with per-list digests. It treats $\{\mathrm{root}_i\}_{i\in[n_{\mathrm{list}}]}$ as witness list-root digests and computes
\[
\mathrm{hash}_i := \mathsf{Hash}(i,\mu_i,\mathrm{root}_i),\qquad i\in[n_{\mathrm{list}}],
\]
then recomputes the top-level root
\[
\mathrm{root}'_{\mathrm{mk}} := \mathsf{MerkleTree}(\mathrm{hash}_0,\dots,\mathrm{hash}_{n_{\mathrm{list}}-1})
\]
and enforces $\mathrm{root}'_{\mathrm{mk}}=\mathrm{root}_{\mathrm{mk}}$ inside the circuit. This pins all centroids and fixes the committed list digests $\{\mathrm{root}_i\}$ without opening any full inverted lists.

\noindent\textbf{Opened probed lists.}
For each probed list $i\in P(q)$ and each slot $j\in[n]$, the circuit hashes the opened record
\[
\mathrm{hash}_{i,j} := \mathsf{Hash}\bigl(i,j,f_{i,j},\mathrm{item}_{i,j},v_{i,j}^{(0)},\dots,v_{i,j}^{(M-1)}\bigr),
\]
recomputes the list root
\[
\mathrm{root}'_i := \mathsf{MerkleTree}(\mathrm{hash}_{i,0},\dots,\mathrm{hash}_{i,n-1}),
\]
and authenticates it to $\mathrm{root}_{\mathrm{mk}}$ at position $i$ via a Merkle opening inside the proof. Concretely, the circuit computes the corresponding top-level leaf digest
\[
\mathrm{hash}'_i := \mathsf{Hash}(i,\mu_i,\mathrm{root}'_i),
\]
and, using the witness sibling digests along the path, recomputes the implied root
\[
\mathrm{root}''_{\mathrm{mk}} := \mathsf{MerkleOpen}(\mathrm{hash}'_i, i; \pi_i),
\]
enforcing $\mathrm{root}''_{\mathrm{mk}}=\mathrm{root}_{\mathrm{mk}}$. Here $\pi_i$ denotes the Merkle authentication path (sibling digests) for leaf position $i$, and $\mathsf{MerkleOpen}(\cdot)$ is the standard iterative hashing procedure determined by the bits of $i$.
Since Step~4 consumes all $n$ slots of each probed list under our fixed-shape semantics, the circuit can recompute $\mathrm{root}'_i$ directly from the opened records; no per-slot Merkle authentication paths are needed.
Together, these constraints ensure that all snapshot values used by the IVF-PQ trace (centroids, PQ codebooks, and the entries in the probed inverted lists) are consistent with the version identifier $\comm$, while keeping the snapshot contents hidden.

\paragraph{Cost}
We instantiate the Merkle circuit using a SNARK-friendly hash (Poseidon in our Plonky2 implementation) and analyze it under a standard hash-cost abstraction. The resulting asymptotic gate complexity is summarized by Equation~\ref{eq:gcommit} and composes additively with the query-circuit costs.

\subsection{Gadgets and Cost Model}
\label{subsec:gadgets}

We express all constraints as arithmetic relations over a field $\mathbb{F}$ and track circuit size via arithmetic gate counts. Witness values are range-bounded to avoid wraparound in comparisons; we write $t_{\mathrm{cmp}}$ for the bit-length parameter used by our range-bounded comparison gadget in the circuit.

Many subcircuits used by \sysname are standard (e.g., fixed-point arithmetic for distance/LUT construction, Poseidon hashing, and Merkle-path verification). Since our main optimization is to replace in-circuit control flow with multiset relations, we highlight the \emph{set-oriented} gadgets that are central to the multiset-based design and to our gate-count analysis.

\paragraph{Tuple compression ($\mathsf{Compress}$)}
We reduce tuple-valued relations to set relations over single field elements via a transcript-derived challenge $\beta\in\mathbb{F}$.
For a tuple $\mathbf{x}=(x_0,\dots,x_{L-1})$, we define
\[
\mathsf{Compress}(\mathbf{x};\beta)=\sum_{i=0}^{L-1}x_i\beta^i,
\]
which can be evaluated via Horner's rule using $L-1$ multiplications and $L-1$ additions, for a total of $2L-2=\Theta(L)$ arithmetic gates.

\paragraph{Range-bounded comparison ($\mathsf{Cmp}$)}
Let $t_{\mathrm{cmp}}=\lfloor \log_2 |\mathbb{F}| \rfloor$ and assume $x,y\in[0,2^{t_{\mathrm{cmp}}-1})$. We form
\[
\Delta=x-y+2^{t_{\mathrm{cmp}}-1}\in(0,2^{t_{\mathrm{cmp}}}).
\]
Enforcing a $t_{\mathrm{cmp}}$-bit decomposition of $\Delta$ yields a boolean output indicating whether $x<y$. The resulting gate cost is $\Theta(t_{\mathrm{cmp}})$.

\paragraph{Multiset equality ($\mathsf{SetEq}$)}
After compressing tuples, multiset equality between two size-$L$ sequences $X=\{x_0,\dots,x_{L-1}\}$ and $Y=\{y_0,\dots,y_{L-1}\}$ can be enforced via a single polynomial identity with a random challenge $\alpha\in\mathbb{F}$:
\[
\prod_{i=0}^{L-1}(\alpha-x_i)=\prod_{i=0}^{L-1}(\alpha-y_i).
\]
Evaluating one side uses $L$ subtractions and $L-1$ multiplications, so evaluating both sides costs $4L-2=\Theta(L)$ arithmetic gates. This captures the essence of permutation-style arguments used in PLONK systems~\cite{plonk}.

\paragraph{Multiset inclusion ($\mathsf{Incl}$)}
To prove multiset inclusion $X\subset_M Y$ with multiplicities, let $n_{\max}=\max\{|X|,|Y|\}$ and pad both multisets to length $n_{\max}$.
The prover supplies: (i) a nondecreasing ordering $X''$ of the padded $X$ and (ii) an aligned sequence $Y''$ whose multiset equals the padded $Y$.
The circuit enforces:
(a) $\mathsf{SetEq}(X,X'')$ and $\mathsf{SetEq}(Y,Y'')$;
(b) nondecreasing order of $X''$ via $(n_{\max}-1)$ comparisons; and
(c) multiplicity alignment via $x''_0=y''_0$ and $(x''_i-x''_{i-1})(x''_i-y''_i)=0$ for all $i\ge 1$.
Overall, this uses two $\mathsf{SetEq}$ relations, $(n_{\max}-1)$ comparisons, and $O(n_{\max})$ alignment constraints. The comparisons dominate, giving an asymptotic cost $\Theta(t_{\mathrm{cmp}}\,n_{\max})$. This gadget is a circuit-level abstraction of lookup-style arguments in PLONK systems~\cite{plookup}.

Table~\ref{tab:gadget-gates} summarizes the asymptotic gate complexity of the above set-oriented gadgets and also lists a few additional primitives used by the baseline circuit-only instantiation; we treat the latter as standard and do not expand their constraints.


\begin{table}[t]
    \centering
    \begin{tabular}{l l l l}
        \toprule
        Gadget & Gate complexity & Gadget & Gate complexity \\
        \midrule
        $\mathsf{Compress}$   & $\Theta(L)$                            & $\mathsf{SetEq}$      & $\Theta(L)$ \\
        $\mathsf{Cmp}$        & $\Theta(t_{\mathrm{cmp}})$             & $\mathsf{Incl}$       & $\Theta(t_{\mathrm{cmp}}\,n_{\max})$ \\
        $\mathsf{BubblePass}$ & $\Theta(L\,t_{\mathrm{cmp}})$          & $\mathsf{LookUp}$     & $\Theta(L)$ \\
        $\mathsf{Permute}$    & $\Theta(L)$                            &                    ---   &       ---       \\
        \bottomrule
    \end{tabular}
    \caption{Core gadgets and their asymptotic arithmetic-gate complexity. Their informal roles are tuple compression, bounded comparison, one pass of a comparison network, swap propagation to a parallel list, multiset equality, multiset inclusion, and table lookup, respectively. Here $L$ is the tuple/list/table length, $t_{\mathrm{cmp}}$ is the comparison bit-length, and $n_{\max}=\max\{|X|,|Y|\}$ for $\mathsf{Incl}$.}
    \label{tab:gadget-gates}
\end{table}

We now use these gadgets to describe two query-circuit instantiations: the circuit-only baseline and our multiset-based design.

\subsection{Baseline Circuit-Only Design}
\label{subsec:baseline-circuit-only}

The circuit-only baseline proves the five-step IVF-PQ semantics by encoding all data-dependent operations \emph{inside} the circuit.
In particular, it realizes the ordering requirements in probe selection (Step~2) and final top-$k$ extraction (Step~5) using explicit comparison networks, and realizes PQ-table access in candidate scoring (Step~4) via in-circuit random access.
Steps~1 and~3 use standard fixed-point arithmetic constraints for distance computations.

\paragraph{In-circuit ordering and selection}
To enforce the Step~2 and Step~5 partial-order conditions, the baseline applies comparison-network passes (e.g., $\mathsf{BubblePass}$) to partially sort the corresponding distance sequences, while applying the same swaps to the associated index/payload sequences via $\mathsf{Permute}$.
This yields a cost of $\Theta(n_{\mathrm{probe}}n_{\mathrm{list}}t_{\mathrm{cmp}})$ gates for probe selection and $\Theta(kn_{\mathrm{probe}}nt_{\mathrm{cmp}})$ gates for the final top-$k$ extraction.

\paragraph{In-circuit PQ-table access}
For each candidate slot and each sub-quantizer, the baseline selects $\mathsf{LUT}_{i,m,v^{(m)}_{i,j}}$ from a length-$K$ table via a $\mathsf{LookUp}$ gadget and sums the selected entries to obtain the approximate distance, with invalid (padding) slots masked to a public upper bound.
These per-candidate table lookups dominate Step~4, contributing $\Theta(n_{\mathrm{probe}}nMK)$ gates.

\paragraph{Gate complexity}
Overall, the baseline query-semantics circuit is dominated by comparison networks (Steps~2 and~5) and in-circuit random access (Step~4), leading to the gate complexity summarized in Equation~\ref{eq:gbaseline}.
Composing it with snapshot binding (Equation~\ref{eq:gcommit}) yields the end-to-end total in Equation~\ref{eq:gtotal-baseline}.
These bottlenecks motivate our multiset-based instantiation below, which proves the same semantics while avoiding in-circuit sorting and random access.

\subsection{Multiset-Based Design}
\label{subsec:multiset-design}

Our multiset-based design removes most data-dependent control flow from the circuit. The prover executes the IVF-PQ pipeline off-circuit to obtain the sorted intermediate sequences and the LUT contributions selected by PQ codes. The circuit recomputes the arithmetic kernels (centroid distances and LUT tables) and then verifies that the prover-supplied trace is consistent with these base quantities, using multiset relations plus lightweight order/boundary checks inside the proof.
This suffices because, in our reference semantics, sorting/selection and random access only serve to justify partial-order conditions and lookup consistency. We therefore take the sorted sequences and selected LUT contributions as witness values and verify them via multiset equality/inclusion relations together with lightweight comparison.

\paragraph{Shared steps (same arithmetic kernels as the baseline)}
Steps~1 and~3 are identical to the baseline: the circuit recomputes all centroid distances $d_i$ and the full ADC tables $\mathsf{LUT}_{i,m,k}$ using fixed-point distance constraints in the same way.
Moreover, the arithmetic part of Step~4 (summing $M$ LUT entries and applying the validity mask with $d_{\max}$) is unchanged; the only difference in Step~4 is how we justify that the prover-selected LUT entries are consistent with the circuit-recomputed tables.

\paragraph{Step 2. Probe selection via multiset equality + boundary checks}
The circuit computes the centroid distances $\{d_i\}$ as in Step~1 and receives a witness sequence:
\[
\begin{aligned}
S^{\mathrm{sorted}}_{\mathrm{idx,dist}}
&= \big((i_0,d_{i_0}),(i_1,d_{i_1}),\dots, (i_{n_{\mathrm{list}}-1},d_{i_{n_{\mathrm{list}}-1}})\big).
\end{aligned}
\]
It then enforces the Step~2 relation by composing gadgets as follows in the circuit.
First, it applies $\mathsf{Compress}$ to each pair $(i_t,d_{i_t})$ (a length-$2$ tuple) and uses $\mathsf{SetEq}$ to prove multiset equality between $S^{\mathrm{sorted}}_{\mathrm{idx,dist}}$ and $S^{\mathrm{orig}}_{\mathrm{idx,dist}}$; since indices in $S^{\mathrm{orig}}_{\mathrm{idx,dist}}$ are unique, this implies that $(i_0,\dots,i_{n_{\mathrm{list}}-1})$ is a permutation of $[n_{\mathrm{list}}]$ and binds each claimed distance to the circuit-computed $d_i$.
Second, it enforces the Step~2 partial-order conditions using $\mathsf{Cmp}$: $(n_{\mathrm{probe}}-1)$ comparisons enforce $d_{i_0}\le\cdots\le d_{i_{n_{\mathrm{probe}}-1}}$, and $(n_{\mathrm{list}}-n_{\mathrm{probe}})$ comparisons enforce $d_{i_{n_{\mathrm{probe}}-1}}\le d_{i_t}$ for all $t\in\{n_{\mathrm{probe}},\dots,n_{\mathrm{list}}-1\}$.
This realizes Step~2 with $\Theta(n_{\mathrm{list}}t_{\mathrm{cmp}})$ comparisons, matching the probe-selection term in Equation~\ref{eq:gmultiset} in our analysis.

\paragraph{Step 4. Candidate distances via multiset inclusion}
The circuit constructs the full ADC lookup tables $\mathsf{LUT}_{i,m,k}$ for all probed lists as in Step~3.
Instead of performing an in-circuit $\mathsf{LookUp}$ for each PQ index, the prover supplies the selected LUT entries
$\ell^{(m)}_{i,j}=\mathsf{LUT}_{i,m,v^{(m)}_{i,j}}$ as witness values.
The circuit proves the correctness of these witness entries by a multiset inclusion check.
After compressing tuples, it enforces that the multiset
\[
\bigl\{(i,m,v^{(m)}_{i,j},\ell^{(m)}_{i,j}) \;\big|\; i\in P(q),\, j\in[n],\, m\in[M]\bigr\}
\]
is included in the multiset of all LUT tuples
\[
\bigl\{(i,m,k,\mathsf{LUT}_{i,m,k}) \;\big|\; i\in P(q),\, m\in[M],\, k\in[K]\bigr\}.
\]
It then computes masked candidate distances $D_{i,j}$ using $d_{\max}$ and the validity flags $f_{i,j}$.
Since the selected-entry multiset has size $|X|=n_{\mathrm{probe}}nM$ while the full LUT multiset has size $|Y|=n_{\mathrm{probe}}MK$, the inclusion gadget operates on $n_{\max}=n_{\mathrm{probe}}M\max\{n,K\}$ elements, yielding the $\Theta(t_{\mathrm{cmp}}n_{\mathrm{probe}}M\max\{n,K\})$ term in Equation~\ref{eq:gmultiset}.

\paragraph{Step 5. Final top-$k$ selection via multiset equality + boundary checks}
Given the flattened candidate list $S^{\mathrm{orig}}_{\mathrm{item,dist}}$, the prover supplies a sorted witness sequence $S^{\mathrm{sorted}}_{\mathrm{item,dist}}$.
The circuit applies $\mathsf{Compress}$ (a length-$2$ tuple) and uses $\mathsf{SetEq}$ to prove multiset equality between the original and sorted candidate sequences.
It then enforces the Step~5 top-$k$ partial-order conditions using $\mathsf{Cmp}$: $(k-1)$ comparisons enforce $D_0\le\cdots\le D_{k-1}$, and $(N_{\mathrm{sel}}-k)$ comparisons enforce $D_{k-1}\le D_t$ for all $t\in\{k,\dots,N_{\mathrm{sel}}-1\}$.
Finally, it outputs the first $k$ payloads in the provided order, which are then certified by the proof.

\paragraph{Gate complexity}
The multiset-based query-circuit complexity is summarized by Equation~\ref{eq:gmultiset}: compared to the baseline, the probe-selection and top-$k$ terms drop from bubble-pass networks to linear-sized boundary checks, and the per-vector LUT lookups are replaced by a single multiset inclusion relation whose size scales with $\max\{K,n\}$ rather than with $nK$ per candidate.
The corresponding end-to-end total complexity is given by Equation~\ref{eq:gtotal-multiset}.

\subsection{Cost-Driven Configuration Selection}
\label{subsec:config_selection}

Our gate-count analysis enables a simple, deployment-oriented configuration-selection procedure.
Unlike plaintext IVF-PQ tuning, which primarily optimizes retrieval utility under latency constraints, \sysname must additionally account for the cost of generating audit proofs.
Proving time is governed by the padded evaluation-domain size $G_B:=2^{\lceil\log_2 G\rceil}$, where $G$ is the circuit gate count; thus, we target minimizing $G_B$ rather than the exact gate count.

\paragraph{Budgets and knobs}
We use the scan budget $N_{\mathrm{sel}}:=n_{\mathrm{probe}}\cdot n$ and the code budget $B:=M\log_2K$.
In our fixed-shape setting, we treat the padded capacity $N:=n_{\mathrm{list}}\cdot n$ and the probing ratio $r:=n_{\mathrm{probe}}/n_{\mathrm{list}}$ as deployment-level choices.
Under a fixed code budget $B$ and power-of-two $K$, the number of sub-quantizers is determined as $M=B/\log_2K$.
Accordingly, our main knobs are $(n_{\mathrm{list}},K)$, which jointly determine $(n_{\mathrm{probe}},n,M)$.

\paragraph{Fixed scan budget: near-linear scaling in $n_{\mathrm{list}}$}
Under fixed $N$ and fixed $r$, we have $n=N/n_{\mathrm{list}}$ and $n_{\mathrm{probe}}=r\,n_{\mathrm{list}}$.
Substituting these relations into the end-to-end cost in Equation~\ref{eq:gtotal-multiset} shows that, for fixed $(B,K)$ and in the common regime $K\le n$, the query-semantics terms are near-linear in $n_{\mathrm{list}}$, while the remaining terms depend only on the fixed scan budget $N_{\mathrm{sel}}$.
Snapshot binding contributes an additive $\Theta(n_{\mathrm{probe}}\log n_{\mathrm{list}})=\Theta(r\,n_{\mathrm{list}}\log n_{\mathrm{list}})$ term due to per-probed-list Merkle openings (Equation~\ref{eq:gcommit}).
In the parameter ranges we study, this sublinear factor has a negligible impact on the observed end-to-end scaling (validated in Experiment~3).

\paragraph{Fixed code budget: a unimodal dependence on $K$}
Under a fixed code budget $B$ (so $M=B/\log_2K$) and fixed scan budget, the $K$-dependent terms in Equation~\ref{eq:gtotal-multiset} can be viewed as a trade-off between:
(i) ADC-table construction, which scales as $\Theta(n_{\mathrm{probe}}KD)$ and increases with $K$; and
(ii) candidate scoring, whose multiset-inclusion component scales as $\Theta(t_{\mathrm{cmp}}\,n_{\mathrm{probe}}M\max\{K,n\})$.
In our regime $K\le n$, the latter simplifies to $\Theta(t_{\mathrm{cmp}}\,N_{\mathrm{sel}}\cdot B/\log_2K)$ and decreases with $K$, implying a unimodal total cost in $K$.
For completeness, when $K>n$ the inclusion term becomes $\Theta(t_{\mathrm{cmp}}\,n_{\mathrm{probe}}B\cdot K/\log_2K)$ and is increasing, so the total cost grows beyond $K=n$.

\paragraph{Bin-pruned search}
To operationalize these observations, we use a simple search procedure that minimizes the padded bin $G_B$ under fixed $(N,B,r)$.
When multiple configurations fall into the same smallest bin, we choose the one with the largest $(n_{\mathrm{list}},K)$ to preserve retrieval utility, since larger $n_{\mathrm{list}}$ (under fixed ratio) and larger $K$ (under fixed $B$) improve recall in plaintext IVF-PQ~\cite{jegou2010product}.

\begin{algorithm}[t]
\caption{Pruned search for a configuration that minimizes the padded bin $G_B$ under fixed code budget $B$ and probing ratio $r$.}\label{alg:bin-pruned-search-fixed}
\begin{algorithmic}
\footnotesize
\REQUIRE Padded capacity $N$, code budget $B$, probing ratio $r:=n_{\mathrm{probe}}/n_{\mathrm{list}}$
\REQUIRE Max $n_{\mathrm{list}}^{\max}$, candidate codebooks $\mathcal{K}$ (powers of two; $M=B/\log_2K\in\mathbb{N}$)
\REQUIRE Gate estimator $\mathsf{GateCount}(D,n_{\mathrm{list}},K)$
\ENSURE Smallest bin $G_B^\star$ and a zk-opt configuration $(n_{\mathrm{list}}^\star,K^\star)$ within that bin

\STATE $N_{\mathrm{sel}} \gets r\cdot N$ \COMMENT{\textcolor{blue}{fixed scan budget}}
\STATE $n_{\mathrm{list}} \gets N/N_{\mathrm{sel}}$ \COMMENT{\textcolor{blue}{minimum feasible layout; then $n_{\mathrm{probe}}=1$}}
\FOR{each $K\in\mathcal{K}$}
  \STATE $G \gets \mathsf{GateCount}(D,n_{\mathrm{list}},K)$
  \STATE $G_B(K) \gets 2^{\lceil \log_2 G\rceil}$
\ENDFOR
\STATE $G_B^\star \gets \min_{K\in\mathcal{K}} G_B(K)$
\STATE $\mathcal{K} \gets \{K\in\mathcal{K}\mid G_B(K)=G_B^\star\}$
\STATE $(n_{\mathrm{list}}^\star,K^\star) \gets (n_{\mathrm{list}}, \max \mathcal{K})$

\WHILE{$\mathcal{K}\neq \varnothing$ \textbf{and} $n_{\mathrm{list}}<n_{\mathrm{list}}^{\max}$}
  \STATE $n_{\mathrm{list}} \gets 2n_{\mathrm{list}}$
  \FOR{each $K\in\mathcal{K}$}
    \STATE $G \gets \mathsf{GateCount}(D,n_{\mathrm{list}},K)$
    \STATE $G_B(K) \gets 2^{\lceil \log_2 G\rceil}$
    \IF{$G_B(K) > G_B^\star$}
      \STATE Remove $K$ from $\mathcal{K}$ \COMMENT{\textcolor{blue}{$K$ leaves the smallest bin}}
    \ENDIF
  \ENDFOR
  \IF{$\mathcal{K}\neq \varnothing$}
    \STATE $(n_{\mathrm{list}}^\star,K^\star) \gets (n_{\mathrm{list}}, \max \mathcal{K})$
  \ENDIF
\ENDWHILE
\RETURN $(G_B^\star,(n_{\mathrm{list}}^\star,K^\star))$
\end{algorithmic}
\end{algorithm}

\section{Experiments}
\label{sec:experiments}

We evaluate \sysname's end-to-end workflow for IVF-PQ top-$k$ retrieval over committed snapshots. Our evaluation covers retrieval utility under ZK-friendly preprocessing, proof-generation cost, and ZK-aware configuration trade-offs.

\paragraph{Evaluation environment}
All experiments are run on a dedicated dual-socket AMD EPYC server with 256 hardware cores and 1~TiB RAM.
We use the default Plonky2 configuration as in the rest of the paper (field, FRI parameters, and recursion layout).

\paragraph{Budgets and bounds}
We use the scan budget $N_{\mathrm{sel}}$ and code budget $B$ to parameterize IVF-PQ configurations.
Prior work suggests that, under these budgets, larger $n_{\mathrm{list}}$ (under a fixed probing ratio) and larger $K$ (under a fixed $B$) often improve recall in plaintext IVF-PQ~\cite{jegou2010product}; in \sysname, we must additionally account for proving cost.
Unless stated otherwise, we fix $B=64$ and the probing ratio $r:=n_{\mathrm{probe}}/n_{\mathrm{list}}=1/128$, which fixes the scan budget once the padded capacity $N$ is set.
We restrict to $n_{\mathrm{list}}\le 8192$ and $K\le 256$.

\paragraph{Evaluation questions}
We evaluate three research questions:
\begin{itemize}[leftmargin=1.5em, topsep=2pt, itemsep=1pt]
    \item \textbf{RQ-(\underline{i}).} \textit{How does ZK-friendly preprocessing affect retrieval utility?} Concretely, after applying fixed-point encoding and fixed-shape rebalancing/padding, how much do accuracy metrics change compared to a floating-point IVF-PQ pipeline?
    \item \textbf{RQ-(\underline{ii}).} \textit{Does our multiset-based design make audit proofs practical?} Concretely, how much does the optimized design reduce gate count and proving time compared to the baseline?
    \item \textbf{RQ-(\underline{iii}).} \textit{Do measured proving costs match our analytic predictions, and does our tuning procedure yield ZK-efficient deployments?} Concretely, under fixed scan/code budgets, do the observed gate counts and proving times follow the trends predicted by our cost analysis, and does Algorithm~\ref{alg:bin-pruned-search-fixed} identify practical zk-opt configurations?
\end{itemize}

\subsection{Retrieval Utility Evaluation}
\label{sec:exp-utility}

Experiment~1 answers Question~(\underline{i}) by evaluating whether our ZK-friendly IVF-PQ preprocessing pipeline preserves retrieval utility.
We evaluate two benchmark families---classic ANN and information retrieval (IR)---and report standard metrics.

\paragraph{Benchmarks and metrics}
We detail the datasets and the corresponding evaluation metrics for each family:
\begin{itemize}[leftmargin=1.5em, topsep=2pt, itemsep=2pt]
  \item \emph{Classic ANN benchmarks (SIFT1M and GIST1M)~\cite{jegou2010product}.}
  We use SIFT1M ($D=128$) and GIST1M ($D=960$) following the original IVF-PQ study.
  For each query, the datasets provide top-100 ordered ground-truth nearest neighbors under $\ell_2$ distance.
  We retrieve the top-$100$ identifiers per query and report Recall@1 and Recall@100, where Recall@$k$ is the fraction of queries whose exact nearest neighbor appears in the top-$k$ retrieved list.
  We also report index construction time and the number of relocated points induced by rebalancing.
  \item \emph{IR benchmark (MS MARCO passage retrieval, dev queries)~\cite{nguyen2016ms}.}
  Each query can have one or more relevant passages; we keep only queries with exactly one relevant passage (which is the dominant case in the dev set) so that each query has a single ground-truth answer passage.
  We embed queries and passages using the sentence embedding model \texttt{msmarco-MiniLM-L6-v3} \cite{reimers2019sentence,wang2020minilm}, yielding $D=384$ vectors, and evaluate ranking quality via Hit@10, MRR@10, and NDCG@10~\cite{craswell2025overview}.
  Since passage retrieval is typically evaluated with cosine similarity, we $\ell_2$-normalize all embeddings before applying our fixed-point encoding, which makes maximizing cosine similarity equivalent to minimizing squared $\ell_2$ distance.
  Here Hit@$k$ is the fraction of queries whose relevant passage appears in the top-$k$ results, MRR@$k$ averages the reciprocal rank truncated at $k$, and NDCG@$k$ is computed with binary relevance and cutoff $k$.
  We retrieve the top-$10$ results per query and compute these metrics on the resulting rankings.
\end{itemize}

\paragraph{Compared systems}
We compare two prototypes:
\begin{itemize}[leftmargin=1.5em, topsep=2pt, itemsep=2pt]
  \item \textit{Std-IVF-PQ (std):} a standard floating-point IVF-PQ pipeline.
  \item \textit{Zk-IVF-PQ (zk):} our ZK-friendly pipeline, which applies fixed-point encoding to all vectors and then runs capacity-constrained rebalancing prior to building the IVF-PQ index.
\end{itemize}
We use the $\ell_2$ distance semantics throughout.
Following our fixed-point representation, we instantiate a 16-bit fixed-point scale in our prototype.
Concretely, we take the maximum absolute coordinate value $v_{\max}$ over all involved real vectors and centroids, rescale it to $2^{16}-1$, and encode each coordinate $v$ as $\lfloor (2^{16}-1)\cdot v / v_{\max} \rceil$ (round-to-nearest integer) before embedding it into the field $\mathbb{F}$.

\paragraph{Configurations}
For each dataset, we test two IVF-PQ settings:
\begin{itemize}[leftmargin=1.5em, topsep=2pt, itemsep=2pt]
  \item \textit{High-acc:} an accuracy-oriented configuration that maximizes $(n_{\mathrm{list}},K)$ within our global bounds ($n_{\mathrm{list}}\le 8192$ and $K\le 256$).
  \item \textit{Zk-opt:} a ZK-oriented configuration selected by Algorithm~\ref{alg:bin-pruned-search-fixed} to minimize the padded bin $G_B$ under the same fixed budgets.
\end{itemize}

\begin{table}[ht!]
  \centering
  \scriptsize
  \setlength{\tabcolsep}{4pt}
  \begin{tabular}{lcccccc}
    \toprule
    Dataset
      & $D$
      & $N_0$
      & $N$
      & $Q$
      & high-acc
      & zk-opt \\
    \midrule
	    SIFT1M
	      & 128
	      & $10^6$
	      & $2^{21}$
	      & $10\,000$
	      & $(8192,64,256)$
	      & $(1024,8,2048)$ \\
	    GIST1M
	      & 960
	      & $10^6$
	      & $2^{21}$
	      & $1\,000$
	      & $(8192,64,256)$
	      & $(512,4,4096)$ \\
	    MS MARCO dev
	      & 384
	      & $8.8\times10^6$
	      & $2^{24}$
	      & $52\,442$
	      & $(8192,64,2048)$
	      & $(2048,16,8192)$ \\
    \bottomrule
  \end{tabular}
  \caption{Experiment~1 datasets and IVF-PQ layouts. $N_0$ is the raw dataset size, $N$ is the capacity after padding, and each layout is reported as $(n_{\mathrm{list}},n_{\mathrm{probe}},n)$ for \textbf{high-acc} and \textbf{zk-opt}. PQ parameters are $(M,K)=(8,256)$ throughout.}
  \label{tab:exp1-setup}
\end{table}

\paragraph{Summary}
Experiment~1 spans three datasets (SIFT1M, GIST1M, and MS MARCO dev) and compares two pipelines (std vs.\ zk) under two configurations (high-acc and zk-opt); Table~\ref{tab:exp1-setup} summarizes the dataset statistics and the corresponding layouts.
Since each dataset is evaluated over a large set of queries, we report metric values computed over the full query set and omit confidence intervals.

\paragraph{Classic ANN benchmarks (SIFT1M and GIST1M)}
Table~\ref{tab:ann-accuracy} reports the results under the Experiment~1 setup (Table~\ref{tab:exp1-setup}). Overall, ZK-friendly preprocessing preserves retrieval utility across both datasets and both configurations: the Recall@1 and Recall@100 values closely track the floating-point baseline, with differences on the order of $10^{-3}$ to $10^{-2}$.
The main cost of the ZK pipeline is offline index construction time, driven by the rebalancing procedure; this overhead is most pronounced for the high-$n_{\mathrm{list}}$ configuration on GIST1M, where more points must be examined and relocated to satisfy the per-list bound.
In all cases, the moved count remains a small fraction of the $10^6$ valid vectors, indicating that the rebalancing perturbation is limited.

\begin{table}[t]
  \centering
  \small
  \setlength{\tabcolsep}{4pt}
  \begin{tabular}{lrrrrrrr}
    \toprule
    Config
      & \multicolumn{2}{c}{R@1}
      & \multicolumn{2}{c}{R@100}
      & \multicolumn{2}{c}{train [s]}
      & zk moved \\
    & std & zk & std & zk & std & zk & count \\
    \cmidrule(lr){2-3}
    \cmidrule(lr){4-5}
    \cmidrule(lr){6-7}
    \midrule
	    \rowcolor{blue!5}\multicolumn{8}{l}{\itshape \textbf{SIFT1M} ($D=128$, $Q=10\,000$)} \\
	      high-acc
	        & 0.5030 & 0.5040 & 0.9527 & 0.9574 & 95.13 & 207.39 & 9\,165 \\
	      zk-opt
	        & 0.4457 & 0.4466 & 0.8719 & 0.8686 & 31.33 & 38.76 & 4\,461 \\
    \midrule
	    \rowcolor{blue!5}\multicolumn{8}{l}{\itshape \textbf{GIST1M} ($D=960$, $Q=1\,000$)} \\
	      high-acc
	        & 0.1903 & 0.1881 & 0.5310 & 0.5230 & 341.56 & 4542.81 & 39\,345 \\
	      zk-opt
	        & 0.1555 & 0.1538 & 0.4140 & 0.4190 & 77.99 & 293.26 & 28\,986 \\
    \bottomrule
  \end{tabular}
  \caption{Utility of std-IVF-PQ (std) vs.\ zk-IVF-PQ (zk) on SIFT1M and GIST1M. We report Recall@1/100, index construction time, and the number of relocated vectors during rebalancing (``zk moved'').}
  \label{tab:ann-accuracy}
\end{table}

\paragraph{IR benchmark (MS MARCO passage retrieval)}
Table~\ref{tab:msmarco-ir} reports the IR metrics under the Experiment~1 setup (Table~\ref{tab:exp1-setup}). Similar to the ANN results, ZK-friendly preprocessing preserves IR utility: Hit@10, MRR@10, and NDCG@10 closely track the floating-point baseline under both configurations.
As expected, the zk-opt configuration trades some accuracy for ZK efficiency compared to the high-acc configuration.

\begin{table}[t]
  \centering
  \scriptsize
  \setlength{\tabcolsep}{3pt}
  \begin{tabular}{lrrrrrrrrr}
    \toprule
    Config
      & \multicolumn{2}{c}{Hit@10}
      & \multicolumn{2}{c}{MRR@10}
      & \multicolumn{2}{c}{NDCG@10}
      & \multicolumn{2}{c}{train [s]}
      & zk moved \\
    & std & zk & std & zk & std & zk & std & zk & count \\
    \cmidrule(lr){2-3}
    \cmidrule(lr){4-5}
    \cmidrule(lr){6-7}
    \cmidrule(lr){8-9}
    \midrule
	    high-acc
	      & 0.3609 & 0.3601 & 0.1914 & 0.1917 & 0.2315 & 0.2316 & 949.91 & 3559.39 & 43\,733 \\
	    zk-opt
	      & 0.3436 & 0.3443 & 0.1807 & 0.1808 & 0.2194 & 0.2197 & 274.38 & 662.77 & 16\,767 \\
    \bottomrule
  \end{tabular}
  \caption{Utility of std-IVF-PQ (std) vs.\ zk-IVF-PQ (zk) on MS MARCO passage retrieval (dev). We report Hit@10, MRR@10, NDCG@10, training time, and relocated-vector count.}
  \label{tab:msmarco-ir}
\end{table}

\begin{tcolorbox}[colback=blue!5!white,colframe=blue!75!black]
\textbf{Answer to RQ-(\underline{i}).}
\textit{ZK-friendly preprocessing largely preserves retrieval utility across our benchmarks: recall- and ranking-based metrics track a floating-point IVF-PQ baseline under both high-acc and zk-opt configurations. The main cost is offline index shaping (rebalancing and padding), which increases index construction time, especially at high $n_{\mathrm{list}}$.}
\end{tcolorbox}

\subsection{Proof Cost Evaluation}
\label{sec:exp-circuit-vs-multiset}

\begin{table*}[t]
  \centering
  \scriptsize
  \setlength{\tabcolsep}{4pt}
  \begin{tabular}{lcccccc}
    \toprule
    System & Prove [s] & Verify [s] & Proof size [kB] & Memory [GiB] & $G$ & $G_B$ \\
    \midrule
    \rowcolor{blue!5}\multicolumn{7}{l}{\itshape \textbf{basic}}\\
    circuit-only     & 2.8547 $\pm$ 0.1067 & 0.0074 $\pm$ 0.0000 & 156.2387 $\pm$ 1.6769 & 12.5341 $\pm$ 0.4172  & 101\,016 & 131\,072($2^{17}$) \\
    multiset-based   & 0.7857 $\pm$ 0.0410 & 0.0076 $\pm$ 0.0017 & 144.2076 $\pm$ 0.8706 & 11.0869 $\pm$ 0.0535  & 21\,198 & 32\,768($2^{15}$) \\
    \midrule
    \rowcolor{blue!5}\multicolumn{7}{l}{\itshape \textbf{low-acc}}\\
    circuit-only     & 0.2696 $\pm$ 0.0114 & 0.0057 $\pm$ 0.0001 & 124.5500 $\pm$ 1.7975 & 12.2770 $\pm$ 0.0000  & 4\,036 & 4\,096($2^{12}$) \\
    multiset-based   & 0.2852 $\pm$ 0.0099 & 0.0058 $\pm$ 0.0001 & 127.9861 $\pm$ 1.1132 & 12.2741 $\pm$ 0.0046  & 5\,929 & 8\,192($2^{13}$) \\
    \midrule
    \rowcolor{blue!5}\multicolumn{7}{l}{\itshape \textbf{large}}\\
    circuit-only     & 527.0096 $\pm$ 51.0197 & 0.0129 $\pm$ 0.0000 & 283.6701 $\pm$ 2.1727 & 747.3327 $\pm$ 7.4720 & 14\,338\,070 & 16\,777\,216($2^{24}$) \\
    multiset-based   & 24.1559  $\pm$ 0.9339  & 0.0108 $\pm$ 0.0000 & 248.9939 $\pm$ 1.1798 & 447.2345 $\pm$ 6.9204 & 915\,627 & 1\,048\,576($2^{20}$) \\
    \bottomrule
  \end{tabular}
  \caption{End-to-end proof cost (including snapshot binding) of the circuit-only baseline vs.\ the multiset-based design under three IVF-PQ configurations (basic, low-acc, and large). Values are mean $\pm$ 95\% confidence interval (5 runs).}
  \label{tab:zkivfpq-bench}
\end{table*}

Experiment~2 answers Question~(\underline{ii}) by benchmarking the circuit-only IVF-PQ baseline against our multiset-based design.
Since verifiability requires binding the proof to a committed snapshot, all results include the snapshot commitment subcircuit.

\paragraph{Configurations}
We consider three representative IVF-PQ configurations and summarize each as a tuple $(N,D,M,K,n_{\mathrm{list}},n_{\mathrm{probe}},k)$, where $N$ is the padded index capacity after rebalancing/padding.
basic is a typical mid-range configuration; low-acc is a deliberately small configuration that favors the circuit-only baseline in gate count; and large is the largest configuration we can run for the circuit-only baseline in our environment (not a limit of the multiset-based design).
\begin{itemize}
  \item \textit{Basic:} $(8192,128,8,16,256,16,64)$;
  \item \textit{Low-acc:} $(8192,128,8,1,16,1,1)$;
  \item \textit{Large:} $(65536,256,16,256,512,64,128)$.
\end{itemize}

\paragraph{Methodology}
We instantiate both designs with identical IVF-PQ parameters and include the same snapshot-binding subcircuit.
For each configuration, we generate and verify proofs 5~times with randomness and report the mean and 95\% confidence interval.

\paragraph{Metrics}
We report end-to-end prove time, verify time, proof size, peak memory (RSS), and circuit size (the number of arithmetic gates $G$ and the padded domain size $G_B:=2^{\lceil \log_2 G\rceil}$, which largely governs proving time due to FFT-based polynomial operations)~\cite{cooley1965algorithm,ben2018fast}.


\paragraph{Results and analysis}
Table~\ref{tab:zkivfpq-bench} shows that in the \textbf{basic} and \textbf{large} settings, the multiset-based design consistently outperforms the circuit-only baseline under snapshot binding.
In particular, it reduces proving time by about $3.6\times$ (basic) and about $22\times$ (large), while keeping verification in the single-digit millisecond range.
In contrast, in the non-practical \textbf{low-acc} setting ($K=1$, $k=1$, $n_{\mathrm{probe}}=1$), the circuit-only baseline can be slightly faster because LUT construction and top-$k$ selection become nearly trivial; the multiset-based design still incurs its set-oriented consistency checks and falls into a larger binned domain size $G_B$.

The $G_B$ column provides a coarse predictor of proving time via the binning effect.
In the remaining experiments, we therefore treat the multiset-based design as the default instantiation of \sysname.

\begin{tcolorbox}[colback=blue!5!white,colframe=blue!75!black]
  \textbf{Answer to  RQ-(\underline{ii}).}
\textit{Yes. For practical IVF-PQ configurations, the multiset-based design makes audit proofs practical under snapshot binding, reducing proving time by up to $\sim 22\times$ while keeping verification in milliseconds.}
\end{tcolorbox}

\subsection{Configuration Trade-offs}
\label{sec:exp-configuration-trade-offs}

Experiment~3 answers Question~(\underline{iii}) by empirically validating the scaling trends predicted by our cost analysis under fixed scan/code budgets, and by applying our tuning procedure to obtain a ZK-efficient deployment configuration used in Experiment~1.

We study SIFT1M ($D=128$) and GIST1M ($D=960$) to cover a wide range of embedding dimensions.
For MS MARCO ($D=384$, $N=2^{24}$), we run only the configuration-selection procedure to obtain the zk-opt configuration used in Experiment~1 and omit the full sweeps for space.

\paragraph{Fixed scan budget varying $n_{\mathrm{list}}$}
We first study how increasing $n_{\mathrm{list}}$ impacts proving cost under a fixed scan budget.
We fix the scan budget by holding the padded capacity $N$ and probing ratio $r:=n_{\mathrm{probe}}/n_{\mathrm{list}}$ constant.
Our analysis predicts that the dominant end-to-end circuit size grows nearly linearly with $n_{\mathrm{list}}$ in this regime, while snapshot binding contributes only a mild additional $\Theta(n_{\mathrm{probe}}\log n_{\mathrm{list}})$ overhead from Merkle openings.

We fix $B=64$ with $(M,K)=(8,256)$ and set $r=1/128$.
For SIFT1M and GIST1M we fix $N=2^{21}$ and sweep $n_{\mathrm{list}}\in\{2^7,2^8,\dots,2^{13}\}$ with $n_{\mathrm{probe}}=n_{\mathrm{list}}/128$ so that $N_{\mathrm{sel}}$ remains fixed.
For each point, we report the end-to-end gate count $G$ and padded evaluation-domain size $G_B$, fit the measured gate counts to a linear function of $n_{\mathrm{list}}$, and benchmark end-to-end proving time to validate that it is primarily governed by $G_B$.
Table~\ref{tab:exp3-nlist-sweep} summarizes the results.

\begin{table}[t]
  \centering
  \small
  \setlength{\tabcolsep}{6pt}
  \begin{tabular}{rrrr}
    \toprule
    $n_{\mathrm{list}}$
      & $T_{\mathrm{prove}}$ [s]
      & $G$
      & $G_B$ \\
    \midrule
    \rowcolor{blue!5}\multicolumn{4}{l}{\itshape \textbf{SIFT1M} ($D=128$)} \\
     128   & $6.7427 \pm 0.7124$  & 182\,225   & 262\,144($2^{18}$)    \\
     256   & $5.7520 \pm 0.1042$  & 192\,846   & 262\,144($2^{18}$)    \\
     512   & $6.0699 \pm 0.0909$  & 213\,475   & 262\,144($2^{18}$)    \\
    1024   & $6.4267 \pm 0.1864$  & 254\,433   & 262\,144($2^{18}$)    \\
    2048   & $12.8939 \pm 0.2739$ & 336\,214   & 524\,288($2^{19}$)    \\
    4096   & $14.1515 \pm 0.2884$ & 499\,721   & 524\,288($2^{19}$)    \\
    8192   & $29.2398 \pm 2.0290$ & 826\,743   & 1\,048\,576($2^{20}$) \\
    \midrule
    \rowcolor{blue!5}\multicolumn{4}{l}{\itshape \textbf{GIST1M} ($D=960$)} \\
     128   & $18.1420 \pm 7.6997$  & 270\,229   & 524\,288($2^{19}$)    \\
     256   & $12.2058 \pm 0.1851$  & 342\,232   & 524\,288($2^{19}$)    \\
     512   & $13.4249 \pm 0.4753$  & 485\,623   & 524\,288($2^{19}$)    \\
    1024   & $28.3921 \pm 4.5745$  & 772\,104   & 1\,048\,576($2^{20}$) \\
    2048   & $61.9585 \pm 11.4705$ & 1\,344\,931 & 2\,097\,152($2^{21}$) \\
    4096   & $124.4228 \pm 25.0629$ & 2\,490\,531 & 4\,194\,304($2^{22}$) \\
    8192   & $263.8778 \pm 45.0721$ & 4\,781\,738 & 8\,388\,608($2^{23}$) \\
    \bottomrule
  \end{tabular}
  \caption{Fixed-scan-budget sweep over $n_{\mathrm{list}}$ for the multiset-based design. We report proving time $T_{\mathrm{prove}}$, end-to-end gate count $G$, and padded bin $G_B$.}
  \label{tab:exp3-nlist-sweep}
\end{table}


A linear fit of $G$ versus $n_{\mathrm{list}}$ across the sweep points yields Pearson $r\ge 0.9999996$ on both datasets, confirming the predicted near-linear scaling under a fixed scan budget.
This supports treating the snapshot-binding term as a low-order deviation in our range.

Proving time exhibits the expected power-of-two ``step'' behavior governed by the padded domain size $G_B$: it changes primarily when the sweep crosses a new $G_B$ bin (e.g., SIFT1M moves from $2^{18}$ to $2^{19}$ at $n_{\mathrm{list}}=2048$ and to $2^{20}$ at $n_{\mathrm{list}}=8192$).
Across the sweep points in Table~\ref{tab:exp3-nlist-sweep}, a least-squares fit of the mean proving time to
$T_{\mathrm{prove}} \approx \alpha\, G_B\log_2 G_B + \beta$,
yields $\alpha \approx 1.36\times 10^{-6}$, $\beta \approx 0.26$, and a Pearson correlation coefficient $r\approx 0.9998$, consistent with the $T=\Theta(G_B\log_2G_B)$ scaling.
We therefore use $G_B$ as the primary lens for configuration selection in the next study.

\paragraph{Fixed code budget varying $K$}
We next study how the PQ codebook size $K$ affects proving cost under a fixed code budget and scan budget.
Our analysis predicts a \emph{unimodal} dependence on $K$: increasing $K$ increases the cost of constructing ADC tables, while decreasing $M=B/\log_2K$ reduces the cost of candidate scoring; in the common regime $K\le n$, this yields a single minimizing $K^\star$.

Since proving time tracks the padded bin $G_B$ (validated above), we perform a gate-only grid search over $K\in\{2,4,16,256\}$ with $M=64/\log_2K$.
For SIFT1M and GIST1M, we set $N=2^{21}$ and evaluate this $K$ grid across $n_{\mathrm{list}}\in\{128,256,512,1024\}$ (with $n_{\mathrm{probe}}=n_{\mathrm{list}}/128$ and $n=N/n_{\mathrm{list}}$), reporting the resulting gate counts and bins.
We also apply Algorithm~\ref{alg:bin-pruned-search-fixed} to MS MARCO under $N=2^{24}$ to obtain its zk-opt configuration used in Experiment~1.
Table~\ref{tab:exp3-gate-grid} reports the grid results under fixed $B=64$ (so $M=64/\log_2K$).

\begin{table}[t]
  \centering
  \small
  \setlength{\tabcolsep}{4pt}
  \begin{tabular}{rcccc}
    \toprule
      & \multicolumn{4}{c}{$K$} \\
    \cmidrule(lr){2-5}
    $n_{\mathrm{list}}$ & 2 & 4 & 16 & 256 \\
    \midrule
    \rowcolor{blue!5}\multicolumn{5}{l}{\itshape \textbf{SIFT1M} ($D=128$)} \\
    128  & 929\,950($2^{20}$) & 497\,482($2^{19}$) & 281\,654($2^{19}$) & 182\,225($2^{18}$) \\
    256  & 935\,409($2^{20}$) & 502\,978($2^{19}$) & 287\,398($2^{19}$) & 192\,846($2^{18}$) \\
    512  & 945\,715($2^{20}$) & 513\,359($2^{19}$) & 298\,272($2^{19}$) & 213\,475($2^{18}$) \\
    1024 & 966\,029($2^{20}$) & 533\,818($2^{20}$) & 319\,722($2^{19}$) & \textbf{254\,433}($2^{18}$) \\
    \midrule
    \rowcolor{blue!5}\multicolumn{5}{l}{\itshape \textbf{GIST1M} ($D=960$)} \\
    128  & 959\,840($2^{20}$) & 527\,830($2^{20}$) & 314\,746($2^{19}$) & 270\,229($2^{19}$) \\
    256  & 994\,980($2^{20}$) & 563\,257($2^{20}$) & 351\,919($2^{19}$) & 342\,232($2^{19}$) \\
    512  & 1\,064\,650($2^{21}$) & 633\,499($2^{20}$) & 425\,652($2^{19}$) & \textbf{485\,623}($2^{19}$) \\
    1024 & 1\,203\,690($2^{21}$) & 773\,683($2^{20}$) & 572\,816($2^{20}$) & 772\,104($2^{20}$) \\
    \bottomrule
  \end{tabular}
  \caption{Gate-count grid over codebook size $K$ under fixed code budget $B=64$ (multiset-based design). Entries report end-to-end gate count $G$ with $G_B$ in parentheses; bold denotes the selected zk-opt configuration.}
  \label{tab:exp3-gate-grid}
\end{table}

Table~\ref{tab:exp3-gate-grid} corroborates our analysis.
For a fixed $K$, gate count increases with $n_{\mathrm{list}}$, and proving time is driven mainly by power-of-two changes in the padded bin $G_B$.
For fixed $(D,n_{\mathrm{list}})$, varying $K$ exhibits a (discrete) unimodal pattern that reflects the trade-off between ADC-table construction and candidate scoring.
When multiple $K$ values fall into the same $G_B$, Algorithm~\ref{alg:bin-pruned-search-fixed} selects the larger $K$ as a utility-preserving tie-breaker, yielding the zk-opt configurations used in Experiment~1.

\begin{tcolorbox}[colback=blue!5!white,colframe=blue!75!black]
  \textbf{Answer to  RQ-(\underline{iii}).}
\textit{Yes. Under a fixed scan budget, measured circuit size scales nearly linearly with $n_{\mathrm{list}}$, and proving time exhibits the expected power-of-two step behavior driven by the padded bin $G_B$. Under a fixed code budget, varying $K$ exhibits the predicted unimodal trade-off. These trends validate our cost model and motivate a cost-driven tuning procedure for selecting zk-efficient configurations.}
\end{tcolorbox}

\section{Related Work}

\paragraph{ANN indices and index shaping}
Vector similarity search is widely deployed, and many ANN indices have been studied at scale~\cite{johnson2019billion}.
We focus on IVF-PQ~\cite{jegou2010product} because its query pipeline admits a standardized, budgeted procedure that can be captured as a fixed-shape semantics.
In contrast, graph-based indices such as HNSW~\cite{malkov2018efficient} rely on adaptive traversals with data-dependent branching and irregular memory access, making ZK authentication substantially more costly.
Our fixed-shape layout is also related to capacity-constrained (balanced) clustering~\cite{ji2025federated,bradley2000constrained,malinen2014balanced,althoff2011balanced,liu2017balanced,liu2018fast,de2023balanced}; we only require a hard per-list bound that enables circuit compatibility.
Unlike prior work that optimizes clustering objectives under such constraints, we use lightweight post-processing to enforce the bound while preserving a IVF-style reference semantics.

\paragraph{ZK proofs for query processing}
ZK proofs have been used to make outsourced query processing publicly verifiable, especially for relational workloads (e.g., vSQL~\cite{vsql}, ZkSQL~\cite{li2023zksql}, and PoneglyphDB~\cite{gu2025poneglyphdb}).
These systems target expressive operators, whereas our workload is a single, highly optimized ANN pipeline.
We target ANN retrieval over high-dimensional vectors, where prover cost is dominated by ordering operations and data-dependent table access, making naive proofs of global scans and full sorting impractical.
\sysname therefore standardizes a fixed-shape IVF-PQ query semantics and designs a proving strategy to this pipeline.

\paragraph{Verifiable nearest-neighbor search and private retrieval}
Prior work studies correctness guarantees for outsourced $k$NN/ANN queries using authenticated indices and verification objects (e.g.,~\cite{wang2024verifiable,cui2023towards}), which can reveal access-dependent metadata and make verification cost scale with the object size.
\sysname targets succinct public verification with ZK protection for embeddings and private index state, using Merkle commitments~\cite{merkle1987digital} instantiated with SNARK-friendly hashes such as Poseidon~\cite{poseidon}.
This avoids exposing index- and access-dependent evidence when the index state is proprietary or sensitive.
Unlike VO-based approaches, the proof can also hide which lists and records are accessed, beyond the fixed circuit shape.
TEEs provide an alternative model for private vector search~\cite{fan2025fedvs,fan2025fedvse}; in contrast, \sysname provides a cryptographic certificate that arbitrary verifiers can check against a public snapshot commitment.
This is particularly relevant when embeddings and ANN index state are proprietary or sensitive (e.g., private RAG corpora~\cite{lewis2020retrieval}).

\section{Conclusion}

We presented \sysname, a verifiable, versioned vector database for audit-on-demand dense retrieval.
An untrusted server answers IVF-PQ top-$k$ queries and, upon challenge, produces a succinct ZK proof that the returned payload list matches the output of a standardized fixed-shape semantics on a publicly committed snapshot, while hiding embeddings and private index state.
To make this practical, \sysname combines ZK-friendly index shaping, a versioned snapshot commitment layer, and a proving backend that avoids in-circuit sorting and random access via multiset checks.
Our Plonky2 prototype achieves up to $\sim 22\times$ faster proving and up to $\sim 40\%$ lower peak memory than a circuit-only baseline while keeping verification in milliseconds, and our experiments show that retrieval utility is preserved and that configurations can be selected using gate-count bins under scan/code budgets.

\bibliographystyle{unsrt}
\bibliography{sample}
\end{document}